# Supported pulmonary surfactant bilayers on silica nanoparticles: Formulation, stability and impact on lung epithelial cells

**F. Mousseau[1]\*, C. Puisney[2], S. Mornet[3], R. Le Borgne[4], A. Vacher[5], M. Airiau[5], A. Baeza-Squiban[2] and J.-F. Berret[1]\***

[1]*Matière et Systèmes Complexes, UMR 7057 CNRS Université Denis Diderot Paris-VII, Bâtiment Condorcet, 10 rue Alice Domon et Léonie Duquet, 75205 Paris, France.*
[2]*Université Paris Diderot – Paris 7, Université Sorbonne Paris Cité, Unité BFA, UMR CNRS 8251, Laboratoire de Réponses Moléculaires et Cellulaires aux Xénobiotiques (RMCX), Paris, France*
[3]*Institut de Chimie de la Matière Condensée de Bordeaux, UPR CNRS 9048, Université Bordeaux 1, 87 Avenue du Docteur A. Schweitzer, Pessac cedex F-33608, France*
[4]*ImagoSeine Electron Microscopy Facility, Institut Jacques Monod, UMR 7592, CNRS - Université Paris Diderot Paris-VII, Paris, France.*
[5]*Solvay Research & Innovation Centre Paris, 52 rue de la Haie Coq, 93306 Aubervilliers cedex*

**Abstract:** Studies have shown that following exposure to particulate matter, the ultrafine fraction (< 100 nm) may deposit along the respiratory tract down to the alveolar region. To assess the effects of nanoparticles in the lungs, it is essential to address the question of their biophysicochemical interaction with the different pulmonary environments, including the lung lining fluids and the epithelia. Here we examine one of these interactive scenarios and study the role of supported lipid bilayers (SLB) on the fate of 40 nm fluorescent silica particles towards living cells. We first study the particle phase behavior in presence of Curosurf®, a pulmonary surfactant substitute used in replacement therapies. It is found that Curosurf® vesicles interact strongly with the nanoparticles, but do not spontaneously form SLBs. To achieve this goal, we use sonication to reshape the vesicular membranes and induce the lipid fusion around the particles. Centrifugal sedimentation and electron microscopy are carried out to determine the optimum coating conditions and layer thickness. We then explore the impact of surfactant SLBs on the cytotoxic potential and interactions towards a malignant epithelial cell line. All *in vitro* assays indicate that SLBs mitigate the particle toxicity and internalization rates. In the cytoplasm, the particle localization is also strongly coating dependent. It is concluded that SLBs profoundly affect cellular interactions and functions *in vitro* and could represent an alternative strategy for particle coating. The current data also shed some light on potential mechanisms pertaining to the particle or pathogen transport through the air-blood barrier.

**Keywords**: Silica nanoparticles – Pulmonary surfactant – Curosurf® – Cryo-electron microscopy – Bio-nano interfaces – Supported lipid bilayer – Cytotoxicity – Cellular uptake
Corresponding authors: jean-francois.berret@univ-paris-diderot.fr, to appear in Nanoscale

# I – Introduction

Airborne materials released by industrial activities are responsible for major heart and lung diseases, affecting an increasing number of humans in both developed and developing countries. Studies have shown that in polluted areas particulate matter can induce various adverse health effects such as premature mortality, asthma, allergic sensitization, and lung cancer.[1-3] In the lungs, it has been found that particles are able to deposit all along the respiratory tract. To evaluate the effects of airways exposure to nanoparticles, it is thus essential to address the question of their biophysicochemical interaction within the different pulmonary environments, from lung lining fluids to epithelia and tissues.

The human respiratory system is composed of channels and ducts where air passes through the airways to reach the pulmonary alveoli during breathing. In the respiratory zone, alveoli provide a large area that favors the transfer of oxygen into the bloodstream and that of carbon dioxide from the bloodstream out of the lungs. With an average size of 200 µm, alveoli are made of thin-walled parenchymal cells (0.5 µm) that are in direct contact with capillaries of the circulatory system.[2] The air-wall interfaces are covered with pulmonary surfactant, which is essential for the lung physiology and also represents the first barrier against pathogens and harmful particles. Depending on their physico-chemical features nanoparticles can eventually cross the alveolar spaces and reach the bloodstream.[1,2,4] Despite continuous efforts, the mechanisms of particles crossing the air-blood barrier are not yet known.

In the search for appropriate systems to assess interactions with nanomaterials, different strategies have been pursued. Some studies focus on endogenous surfactant that is isolated from porcine broncho-alveolar lavage fluid.[2,4-10] In such cases, the phospholipid dispersion contains the four main proteins SP-A, SP-B, SP-C and SP-D associated with pulmonary surfactant and their relative proportions are close to those *in vivo*. A second approach takes advantage from existing exogenous surfactants administered to premature infants of less than 32 weeks.[9,11-21] These surfactant formulations are of porcine (Curosurf®, Survanta®) or bovine (Alveofact®) origin and are part of efficient surfactant replacement therapies. Because they possess both SP-B and SP-C hydrophobic proteins, these substitutes are considered as reliable surfactant models. Finally, bilayers made from synthetic phospholipids are also used to mimic the physico-chemical interactions with particles.[6,7,16,22-35] These dispersions usually consist of one or two types of lipids and do not contain any of the above-cited proteins. Particle interaction studies with pulmonary surfactants have shown a wide range of behaviors, including the formation of aggregates, decorated vesicles, supported lipid bilayers or of particles internalized inside the membrane.[27,36-38] Working with Curosurf® at physiological concentration, Schleh *et al.* was one of the first to study the effect of nano- and microparticles on the structural and interfacial properties of the lung fluid.[17] The authors found that bare titanium dioxides strongly associate with surfactant lipid membranes and modify their surface tension and layered structures. At non-physiological concentrations, several studies reported the formation of large aggregates (> 1 µm) or more generally an amelioration of particle stability in presence of surfactant.[16,20,39]

Here we focus on supported lipid bilayers (SLBs) deposited on nanoparticles. In real-life situations, particles crossing the air-liquid interface in the alveolar spaces may indeed be covered with SLBs.[2,5] In this context, we aim at providing new understanding on the role of biomolecular films at nano-interfaces and their impact *in vitro*. Supported lipid bilayers were originally produced on planar surfaces using Langmuir-Blodgett/Langmuir-Schaefer deposition techniques, or by vesicle fusion. The mechanism of SLB formation on flat substrates is now well understood and results from a balance between hydration, Van der Waals and electrostatic dispersion forces.[40,41] On highly curved interfaces such as exhibited by nanoparticles however there has been less research.[42] To make SLB-coated particles, several routes were explored and the lipid deposition was mediated using different approaches including liposomal fusion, solvent exchange, ionic strength variation or sonication.[36,42] In recent years, SLBs have been obtained in combination with metal oxide,[25,27,29,32,34,35] noble metal particles[37] and with polymeric nanogels.[13,14] However with a few exceptions,[12,14] the SLB deposition on highly curved surfaces was obtained mainly using zwitterionic lipids of synthetic origin, and not with biological lipid mixtures.[27,42]

In this work we propose a simple method to coat 40 nm model fluorescent silica nanoparticles with a supported bilayer, the bilayer being that of the exogenous surfactant Curosurf®. To our knowledge, it is the first time that SLBs from surfactant substitute are achieved on particles below 100 nm. Silica particles are chosen because their physico-chemical features (size, surface charge, optical properties) can be accurately tuned by synthesis. For imaging and visualization purposes, the particles were made fluorescent by incorporating rhodamine molecules in the silica structure. To enhance the particle-vesicle adhesion energy and favor SLB formation *via* Coulombic forces,[43,44] the particles are coated with high-density amine groups. At first we examine the particle-vesicle phase behavior as a function of concentration and temperature. Such an approach provides critical information on the colloidal interactions and assembled structures. It is shown that the nanoparticle-vesicle association is mainly driven by electrostatics and leads to large (> 1 µm) aggregates. The aggregates are found to be stable over time and the membranes do not reverse spontaneously into SLBs. To produce SLBs, we use sonication to reshape the vesicles and induce the lipid fusion around the particles. The vesicle-to-particle specific surface is optimized to have all silica particles coated with a lipid bilayer. We also address the question whether SLBs affect the particle interaction and uptake towards alveolar epithelial cells, yielding a positive response: SLB-coated particles exhibit a $20 - 50$ times lower internalization compared to bare particles. We finally propose a mechanistic scenario to explain the reduction of particle toxicity and uptake, and suggest that in case of SLB formation, pulmonary surfactant tends to alleviate the negative impact of nanomaterials in alveolar spaces.

# II – Results and discussion

## II.1 – Nano supported lipid bilayer formulation

Figs. 1a and 1b display representative cryo-TEM images of Curosurf® diluted at 5 g L⁻¹. Images show that the phospholipids self-assemble locally into a bilayer structure of thickness $\delta = 4.36 \pm 0.39$ nm (Supporting Information S1). When observed on a larger scale, the bilayers are found to close on themselves and form unilamellar and multivesicular vesicles. Multivesicular vesicles describe large compartments encapsulating one or several smaller vesicles.[45] The vesicle size distribution derived from cryo-TEM extends from 50 to 1000 nm with a median value of 230 nm and a dispersity of 0.55 (Fig. 1c). The phospholipid structures observed for Curosurf® (Figs. 1a-b and S2) are similar to those reported in the literature using cryo- and freeze-fracture electron microscopy.[10,14,46,47] They are also comparable to those of endogenous surfactant obtained from bronchoalveolar lavages.[6] Fig. 1d shows a cryo-TEM image of the fluorescent silica synthesized by the Stöber technique. The particles are nearly spherical and characterized by a diameter of 41.2 nm and a dispersity of 0.11 (Fig. 1e). A particle identity card displaying UV-Vis spectrometry, fluorimetry and light scattering data is provided in Supporting Information (S3). Concerning the surface charges, the particles are positive at pH 6 and below (zeta potential $\zeta = +$ 47 mV) with a charge density of $+0.62e$ nm$^{-2}$ and the vesicles are negative ($\zeta = -55$ mV).[16,48] It is thus anticipated that vesicles and particles will interact strongly upon mixing.[49,50]

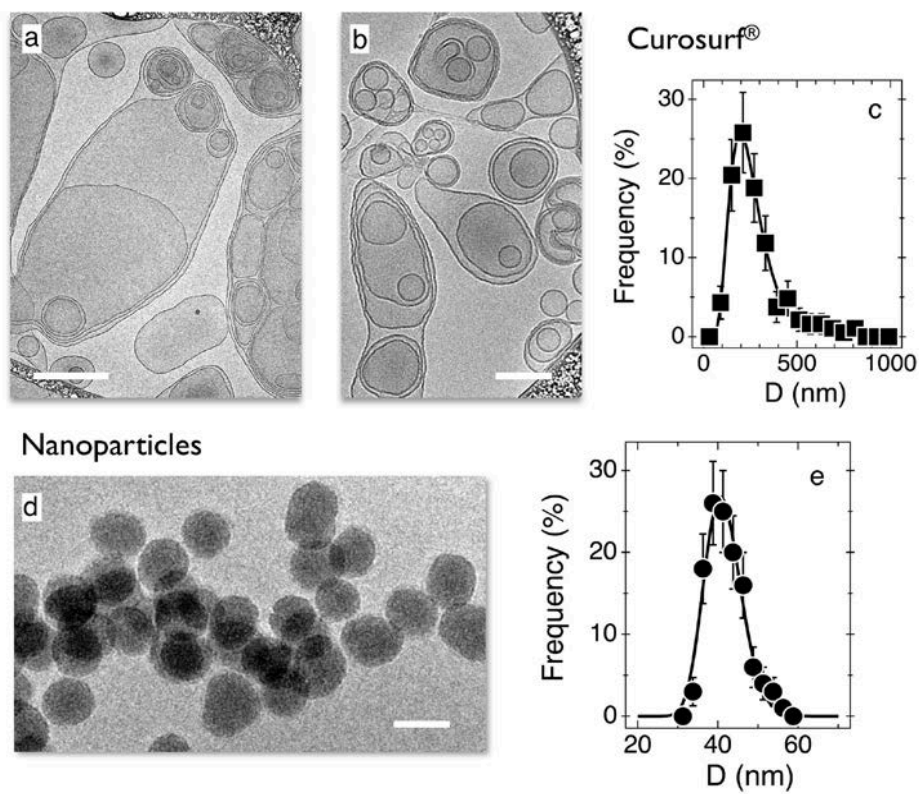

**Figure 1:** *Cryo-TEM images of Curosurf® (a,b) and of rhodamine-loaded fluorescent silica nanoparticles (d). c,e) Size distributions of vesicles and nanoparticles, respectively. For the vesicles, the size of each of the enclosed structures have been included into the full statistical analysis. The distribution is well accounted for by a log-normal function of median 230 nm and dispersity 0.55. For the particles, the median value is 41.2 nm and the dispersity 0.11. Extra Curosurf® illustrations are provided in S2.*

Mixed silica-surfactant dispersions were prepared at different concentration $c$ and surface ratio $X_S$, where $X_S$ denotes the ratio between the specific surface associated with the vesicles and that associated with the particles. With these notations, $X_S = 0$ corresponds to a nanoparticle solution and $X_S = \infty$ to a vesicular dispersion. Detailed calculations for $X_S$ are provided in the Supplementary Section S4. In a first attempt, a protocol described for synthetic lipids was reproduced using Curosurf®.[29] The fluorescent silica particles were mixed with vesicles at pH 6 and concentration 0.1 g L$^{-1}$ with a large excess of membranes ($X_S = 15$). The choice of this pH was imposed by the particle and vesicle physico-chemical properties (both are of oppositely charges and stable) and by the fact that it corresponds to the mildly acid conditions encountered in alveolar spaces.[51,52] The mixed dispersion was studied over 24 h by light scattering, and revealed the formation of stable micron-sized aggregates. In a second attempt silica-surfactant mixtures were prepared over a wide range of surface ratios ($X_S = 10^{-4}$ to $10^4$) following the Continuous Variation Method developed by us to study multicomponent systems.[48,50,53] The screening of the phase diagram led to a marked scattering peak centered around $X_S = 2.6$, indicating again the formation of mixed silica-vesicle aggregates over a broad range of conditions (S5).[48] The results were similar for lipids in the gel or in the fluid membrane phase, which transition was determined by differential scanning calorimetry at 29.5 °C (Supplementary Information S6). To check whether the aggregate formation is specific to Curosurf®, additional experiments were performed using a protein-free surfactant constituted from dipalmitoylphosphatidylcholine (DPPC) and from two other lipids.[16] Details about the lipids and their formulation are provided in the Materials and Methods section. The lipid dispersion was extruded through a 100 nm pore polycarbonate membrane, resulting in a highly uniform population of negatively charged vesicles ($\zeta = -30$ mV). Supplementary Information S7 illustrates the static and dynamic light scattering for silica-vesicle mixed dispersions. At the two temperatures investigated (T = 25 °C and 37 °C) it was found that with synthetic lipids sub-micron aggregates formed upon mixing and that the aggregate formation is not specific to the biologically relevant Curosurf® membranes. In conclusion, we show here that mixing positive fluorescent silica and negative Curosurf® vesicles does not lead to the spontaneous formation of supported lipid bilayers, but rather to uncontrolled electrostatics-driven aggregation.[14] The discrepancy with the literature could come from the nature of the phospholipids and their interaction with silica surface groups. Most studies showing a membrane fusion on silica nano- and microparticles were performed using zwitterionic lipids of synthetic origin.[31,32,36,42]

Sonication is a well-known technique to improve particle dispersability. For lipid membranes, sonication helps reshaping the lamellarity through the disruption and reformation of the assembled structures.[54] Fig. 2a illustrates schematically the individual structures, the solid fluorescent silica particles at $X_S = 0$ and the multivesicular vesicles at $X_S = \infty$. Fig. 2b shows the hydrodynamic diameter of a 1 g L$^{-1}$ Curosurf® dispersion ($X_S = \infty$) as a function of the sonication time. $D_H$ decreases rapidly and stabilizes around 72 nm (polydispersity index pdi 0.34) after 20 minutes. Sonication performed on a mixed surfactant sample ($X_S = 2.6$) displays a similar behavior, the diameter reaching a value of 84 nm (pdi 0.31) at steady state. These decreases are in both cases accompanied with a strong reduction in the sample turbidity. Sonication was then carried out for 90 minutes at different $X_S$ and the solutions were investigated *via* light scattering and zetametry. Figs. 2c and 2d show the $D_H(X_S)$ and $\zeta(X_S)$-dependencies for the sonicated dispersions. It is found that above a surface coverage of 2, both quantities exhibit stationary values. Similar results in DI-water and in cell culture medium (DMEM) confirm the robustness in the hybrid structures. Besides, the zeta potential decreases from + 47 mV for the silica particles, passes through zero at $X_S = 1$ and stabilizes at -30 mV above. The charge reversal observed here, together with a hydrodynamic diameter slightly larger than that of the bare silica suggests that for $X_S \geq 2$ the particles are coated with a supported lipid bilayer. The fact that the optimum ratio was not $X_S = 1$ could be due to the assumptions made to evaluate the particle and vesicle specific surfaces (S4).[22] To demonstrate the presence of a double layer at the particle surface and measure its thickness, centrifugal sedimentation (DCS), transmission and cryo-transmission electron microscopy experiments were conducted.

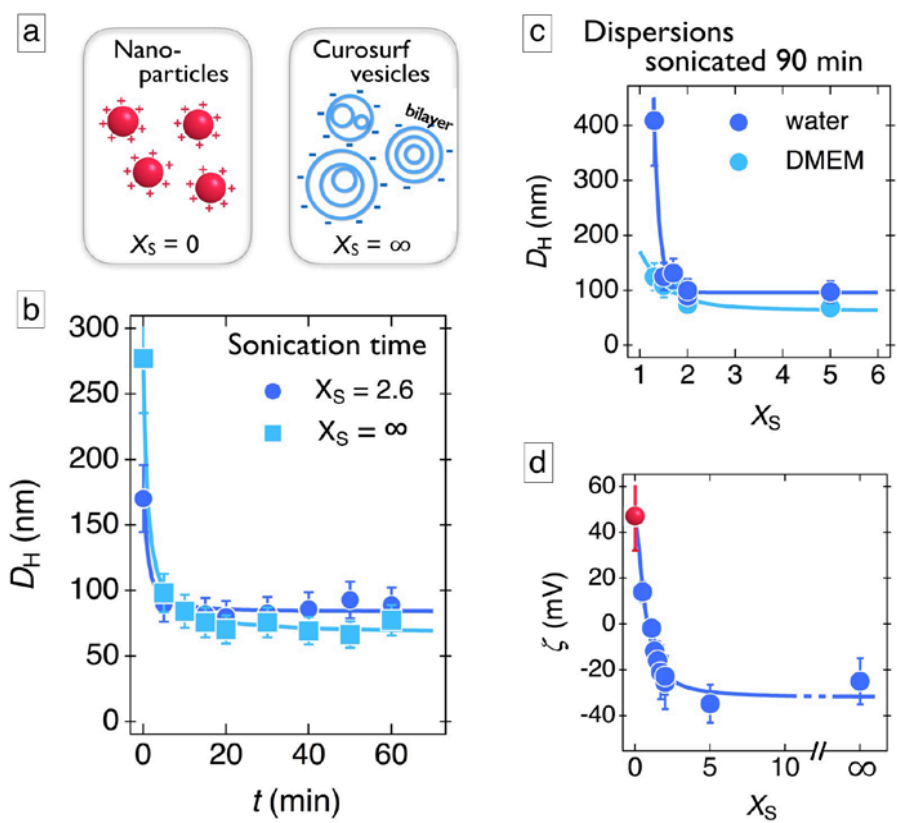

**Figure 2:** *a) Schematic representation of rhodamine-loaded fluorescent silica nanoparticles and of Curosurf® vesicles. The particles at pH 6 and below are positive ($\zeta = + 47$ mV) and the vesicles are negative ($\zeta = - 55$ mV). b) Hydrodynamic diameter $D_H$ versus sonication time for Curosurf® ($X_S = \infty$) and Curosurf® mixed with fluorescent silica nanoparticles ($X_S = 2.6$) at 1 g $L^{-1}$. c and d) Hydrodynamic diameter $D_H$ and zeta potential $\zeta$ as a function of the surface ratio $X_S$ in DI-water and in DMEM cell culture medium.*

## II.2 – Supported bilayer thickness determination

*Differential Centrifugal Sedimentation*

DCS was performed on 1 g $L^{-1}$ sonicated dispersions at various surface ratios $X_S = 0$, 1.3, 1.5, 1.7, 2 and 5. Fig. 3a displays the absorbance measured as the particles are passing through the beam detector and plotted as a function of particle size. In the current configuration the dispersion is injected from the disk center and the sedimentation occurs to the outer range. As a result, mulivesicular vesicles alone cannot be detected, their density being lower than that of the sucrose gradient (1.064 g $cm^{-3}$). For bare silica, the absorbance exhibits a peak centered at 45.0 nm. In presence of surfactant a peak is also observed but its position is shifted to lower values (43.3 nm). As shown in the inset, the data for $X_S = 1.3 - 5$ are well superimposed near to the maximum. In the range $60 - 100$ nm, the absorbance exhibits an additional contribution (arrow). This contribution has the form of a shoulder and results from particle-vesicle aggregates formed through electrostatic interaction, as discussed in the previous section. The DCS data superimposition suggests that $X_S = 2$ leads to a complete particle coverage whilst limiting the excess of surfactant.

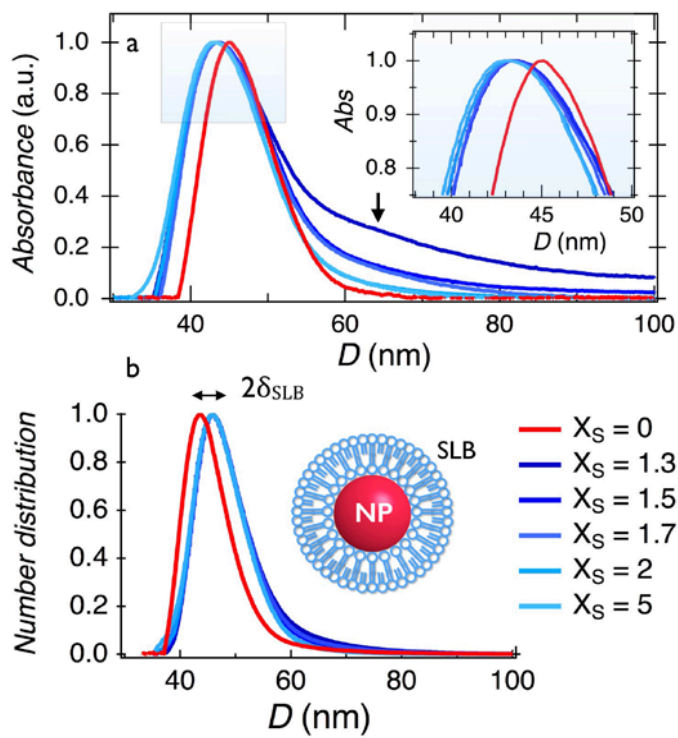

***Figure 3:*** *a) DCS absorbance for 1 g $L^{-1}$ sonicated fluorescent silica-surfactant dispersions at various surface ratios $X_S = 0$, 1.3, 1.5, 1.7, 2 and 5. b) Corresponding number distributions calculated within the single adsorbed bilayer model. Inset in b): Representation of a rhodamine-loaded silica particle coated with a single supported bilayer.*

The shift towards smaller diameter found with Curosurf® (45.0 nm *versus* 43.3 nm for the absorbance and 43.6 nm *versus* 40.8 nm for the size distribution) comes from the fact that the density entered in the DCS software is that of silica ($\rho = 1.9$ g cm$^{-3}$), the one corresponding to actual silica-vesicle structures being unknown. In practice, the mass density indicated leads to an incorrect relationship between the sedimentation time and the size. Similar results were obtained by Krpetic *et al.* with gold nanoparticles and poly(ethylene glycol) coatings.[55] In Supporting Information S8, the Stokes equation relating the sedimentation time $t$, the particle size $D$ and density $\rho$, $(\rho - \rho_0)D^2 t = cste\ \rho$ was evaluated assuming that the particles were coated with a single lipid bilayer of mass density $\rho_{SLB} = 0.9$ g cm$^{-3}$.[56,57] Solving the Stokes equation gives a total diameter of 47.2 nm and a layer thickness of 1.8 nm, which is smaller than the value obtained by cryo-TEM (S1). Fig. 3b displays the number distribution retrieved from the absorbance data at various $X_S$ and corrected from the previous adjustment. In conclusion, DCS is an accurate technique for retrieving both particle size and particle distribution, however in cases of complex colloids a structural model is required *a priori* and the determination is only approximate.

*Transmission electron microscopy (TEM)*

For TEM experiments, 3 μL of a $X_S = 2$ and $c = 0.05$ g L$^{-1}$ silica-surfactant dispersion were deposited on a copper grid for 10 min followed by a subsequent addition of uranyl acetate. This compound provides a dark staining of the phospholipid heads due to the high affinity of uranium oxycation to carboxyl groups. Figs. 4a-c displayed images of the fluorescent silica particles at different scales. The particles are found to be non-aggregated and surrounded with a white coat that is identified as a supported lipid bilayer (blue arrows).[16,47,58,59] A size analysis performed on $n = 200$ leads to the distributions in Fig. 4d. Mean diameters are 41.1 ± 4.6 nm and 56.7 ± 7.2 nm for the bare and SLB-coated particles, resulting in a thickness $\delta_{SLB}$ of 7.2 nm. This value exceeds the bilayer thickness found for phospholipids and for Curosurf®.[60] This discrepancy is attributed to the sample dehydration that leads to irregular contours and increases uncertainty in the SLB thickness determination.[29,47]

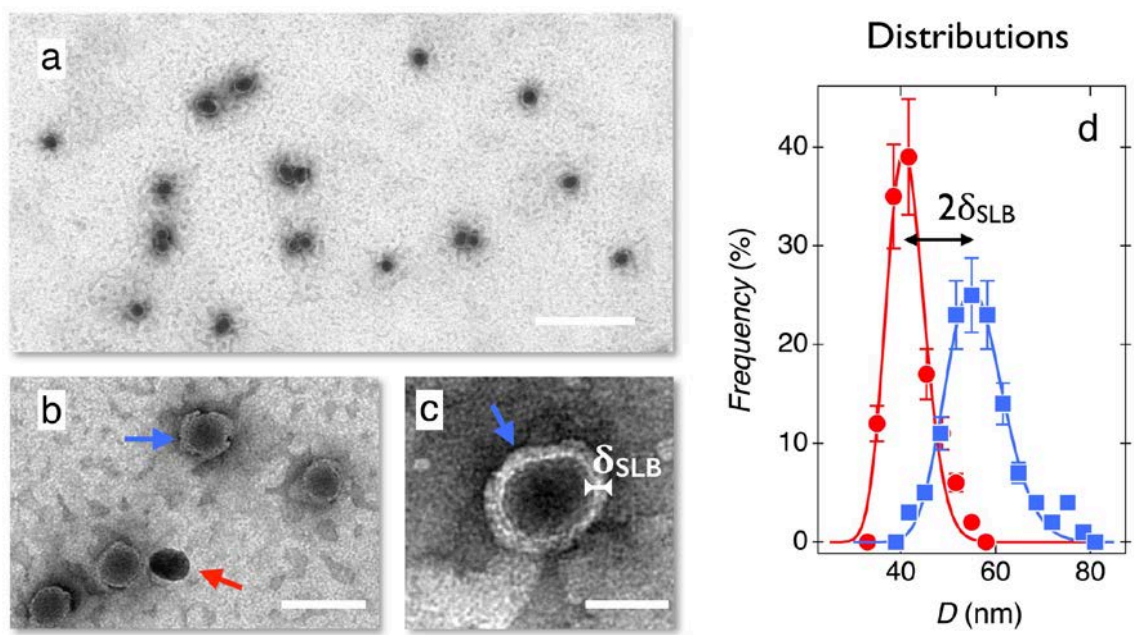

**Figure 4:** *a – c) TEM images of SLB-coated particles. The particles were prepared at 1 g L$^{-1}$ as in Figs. 2 and 3 and diluted down to 0.05 g L$^{-1}$ for observation. The supported lipid bilayer appears as a bright layer thanks to uranyl acetate staining. The bars are 300, 100 and 40 nm respectively. d) Size distributions determined from TEM for bare (red circles) and SLB-coated fluorescent (blue squares) silica. As indicated in the text and in Table I, the bilayer thickness was estimated at 7.2 nm.*

*Cryo-transmission electron microscopy*

Fig. 5a shows an overview of a sonicated silica–surfactant dispersion prepared at the surface ratio $X_S = 2$ ($c = 0.5$ g L$^{-1}$). The image displays a large number of non-aggregated particles coexisting with few vesicles. A statistical analysis ($n = 800$) reveals that less than 5% are indeed vesicles and that their sizes are around 100 nm. Figs. 5b – 5d exhibit close-up views of the particles, which present a thin and bright layer of thickness $5.2 \pm 0.6$ nm at the periphery. These cryo-TEM images, together with others in Supporting Information S9 indicate that every detected particle appear to be decorated with a surfactant outer shell. They also confirm the DCS outcome stating that particle and vesicle specific surfaces should be equivalent around $X_S = 2$. The SLB-coated particles observed by cryo-TEM are similar to those reported with phospholipid formulations.[25,27,29,32,33] and with Curosurf®.[13,14] Table I sums up the different results obtained from light scattering, DCS, TEM and cryo-TEM. Cryo-TEM yields a value that is the closest from that of the membrane, $\delta = 4.36 \pm 0.39$ nm (S1).

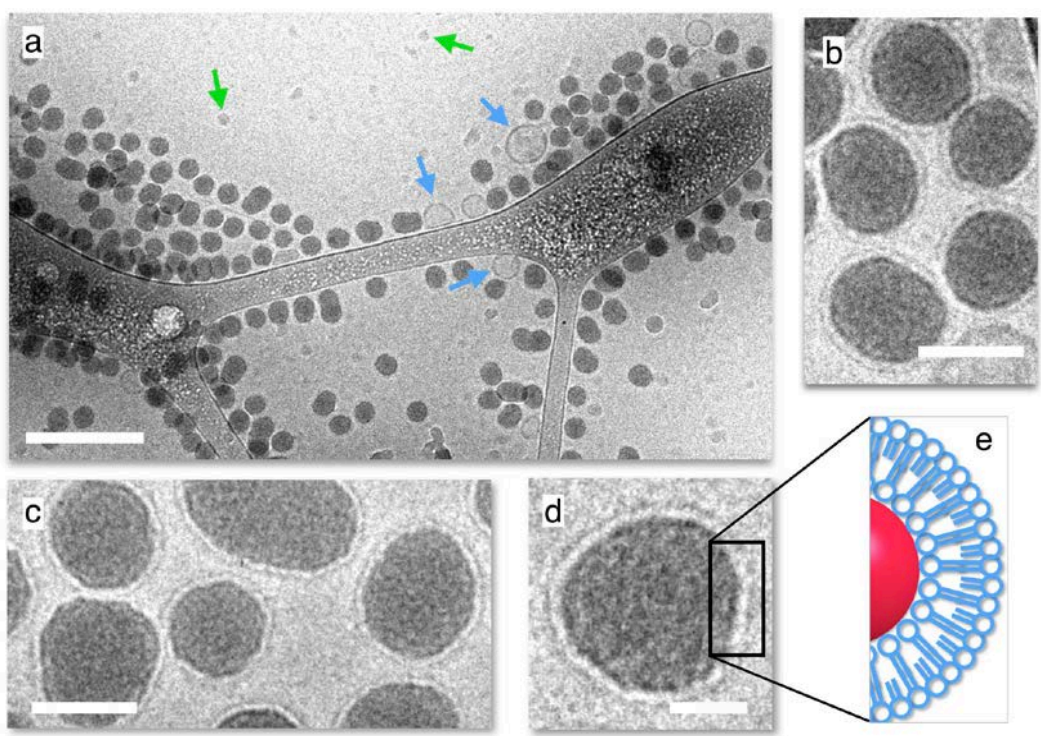

**Figure 5:** *a) Cryo-TEM image of a sonicated fluorescent silica–Curosurf® dispersion ($X_S = 2$). The particles were prepared at 1 g L$^{-1}$ as in Figs. 2-4 and diluted down to 0.5 g L$^{-1}$ for cryo-TEM observation. The blue and green arrows are pointing to surfactant vesicles and ice crystals respectively. The bar is 200 nm. b – d) Close-up views of fluorescent silica nanoparticles coated with a supported lipid bilayer at different magnifications (bars are 40 nm in b and c and 20 nm in d). e) Schematic representation of a SLB coat.*

|  | DLS | DCS | TEM | Cryo-TEM |
|---|---|---|---|---|
| NP diameter (nm) | 60 | 43.6 | 41.1 | 41.2 |
| SLB-coated silica diameter (nm) | 84 | 47.2 | 55.6 | 51.6 |
| $\delta_{SLB}$ (nm) | 12 | 1.8 | 7.2 | 5.2 |

**Table I:** *Fluorescent silica nanoparticle diameter with and without supported lipid bilayer determined by dynamic light scattering (DLS), differential centrifugal sedimentation (DCS), transmission electron microscopy (TEM) and cryogenic transmission electron microscopy (cryo-TEM). The SLB thickness $\delta_{SLB}$ in the last line should be compared to the Curosurf® membrane thickness $\delta = 4.36 \pm 0.39$ nm.*

The colloidal stability of the coated particles was investigated over time and in different solvents. Fig. S10 in Supporting Information shows the time dependence of the hydrodynamic diameter and zeta potential in DI-water over 21 days and in DMEM culture medium with serum over one week. The size and charge were found stationary over the measuring periods, indicating that the supported bilayer remains firmly adsorbed at the particle surface, and that the coating insures long-term electrostatic as well as steric stability. To assess potential interactions with the surfactant vesicles, the Continuous Variation Method was applied.[16,53] Fig. S11 displays the scattered intensity as a function of mixing ratio $X$ between $10^{-3}$ and $10^3$. The scattering data were found to vary continuously from one stock solution to the other, showing indeed a quite different behavior than that of bare aminated silica (S5). In this case, the intensity was well accounted for by the non-interacting model, demonstrating that coated particles and vesicles do not mutually interact. In conclusion, we have shown that SLBs can be prepared on aminated silica using a surfactant substitute. Although particles and vesicles interact strongly *via* electrostatic attraction, SLBs do not form spontaneously upon mixing and need to be assisted by sonication. The SLB-coated fluorescent silica are stable over time and can be used as such for *in vitro* studies.

### II.3 – Role of the SLB towards toxicity and internalization rates

*Toxicity assays*

We hereafter illustrate the effects of SLB coating on the toxicity and on the cell internalization *in vitro*. The cytotoxicity of the bare aminated fluorescent silica and surfactant formulations was assessed using A549 lung epithelial carcinoma cells combined with the WST-1 assay. WST-1 was performed at increasing silica doses from 0.2 to 1000 µg mL$^{-1}$ in serum free medium. Figs. 6 a-c display the percentage of viable cells after a 24 h treatment with bare and SLB-coated silica as well as with Curosurf® alone. For bare particles, the viability exhibits a 50% increase at low dose, remains stationary until 100 µg mL$^{-1}$ and decreases above, indicating a marked toxicity. This increase of viability at low concentration is typical of a hormesis effect related to cell adaptation.[61] Studies have shown that particles with positive surface charges are more harmful than neutral or negative ones, as they are susceptible to cling to the plasma membrane *via* electrostatic interactions and modify locally its integrity. Besides, the effect might be here amplified by the presence of amine groups that are known to induce membrane perforation.[62] In contrast, exposing the A549 cells to surfactant and to SLB-coated fluorescent silica did not cause any significant reduction in cell viability. Fig. 6b and 6c display rather uniform cellular responses with a viability remaining around 100%.

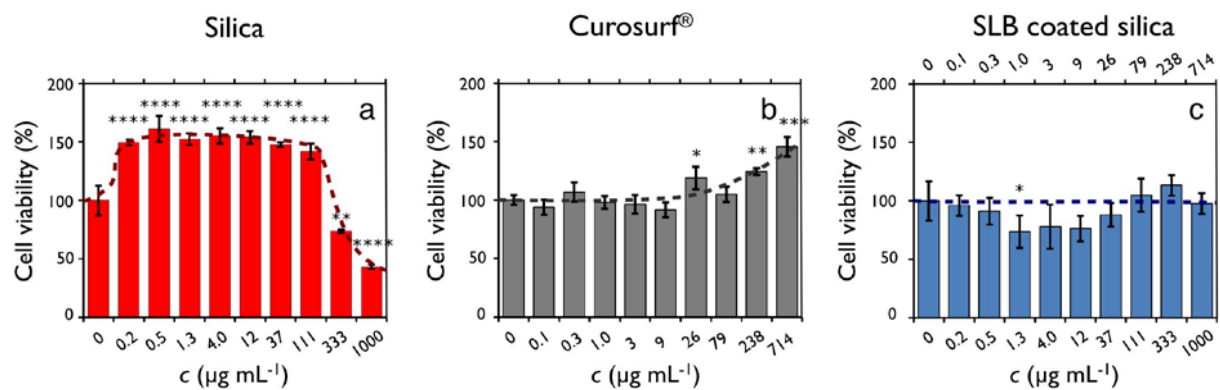

**Figure 6:** *Cell viability of adenocarcinomic human alveolar epithelial A549 cells treated with bare aminated fluorescent silica (a), Curosurf® (b) and with SLB-coated particles (c). In c) the lower scale is that of the particles and the upper scale that of Curosurf®. Mixed nanoparticle-vesicle dispersions in c) were prepared at $X_S = 2$ and sonicated 90 min. Viability data are means ± standards deviations (n = 3) and analyzed using one-way ANOVA and Dunnett's tests with p = 0.0332, 0.0021, 0.002 and 0.0001, corresponding to one to four stars respectively.*

*Confocal microscopy and internalization rates*

For confocal microscopy, the A549 cells were incubated with the rhodamine-loaded fluorescent silica particles at 1 g L$^{-1}$ for 4 and 24 h. Fig. 7A – 7B display the 4 h bright field (first row) and confocal (middle row) images acquired for bare and SLB-coated nanoparticles. The blue and red fluorescence channels correspond to DAPI and rhodamine emissions, respectively. For uncoated fluorescent silica in Fig. 7A (panels a,b) the bright field image reveals the presence of intracellular vacuoles.[63] These vacuoles are not present in cells incubated with SLB-coated particles (Fig. 7B). On fluorescence images, the particles are outlined as dots for both treatments, the dots being less numerous for coated particles. For the bare fluorescent silica, the dots also form large patches suggesting intracellular aggregation (S12). In Fig. 7A and 7B, the particles close to the nuclei are inside the cytoplasm as cells display red fluorescence at nucleus height levels, whereas those at the periphery are either internalized or adsorbed at the plasma membrane. Confocal images at 4 h and 24 h were globally similar.

To provide quantitative figures, a fluorescence-based protocol was developed to evaluate the mass of internalized and adsorbed particles (Fig. 7C). Cells were incubated with the rhodamine-loaded fluorescent silica particles at concentrations 5, 25, 50, 75 and 100 µg mL$^{-1}$ for 4 h using 96-well plates. A reference signal was obtained from bare silica dispersed in white DMEM. Varying linearly with concentration (continuous line in Fig. 7D), this signal allowed translating the fluorescence intensity into masses of internalized and adsorbed particles. Fig. 7D compares the 4 h incubation data for bare aminated and SLB-coated fluorescent silica. In both assays, the experimental mass increases roughly linearly with the applied dose. However, bare silica levels are systematically above those with a SLB coating by a factor 20 to 50. To account for such behaviors, we assume that electrostatics is playing a crucial role. The SLB-coated particles being negatively charged and disperse (S9), they are able to diffuse rapidly and interact with the membrane by random collisions in single particle events. Bare aminated silica in contrast are positively charged and adsorb at the plasma membrane *via* charge complexation. As particles agglomerate in cell medium, mixed aggregates are susceptible to settle and enter in contact with the cell membranes. In conclusion, we have seen that supported lipid bilayers from pulmonary surfactant profoundly affect cellular interactions, probably due to a combination of charge and dispersability effects.

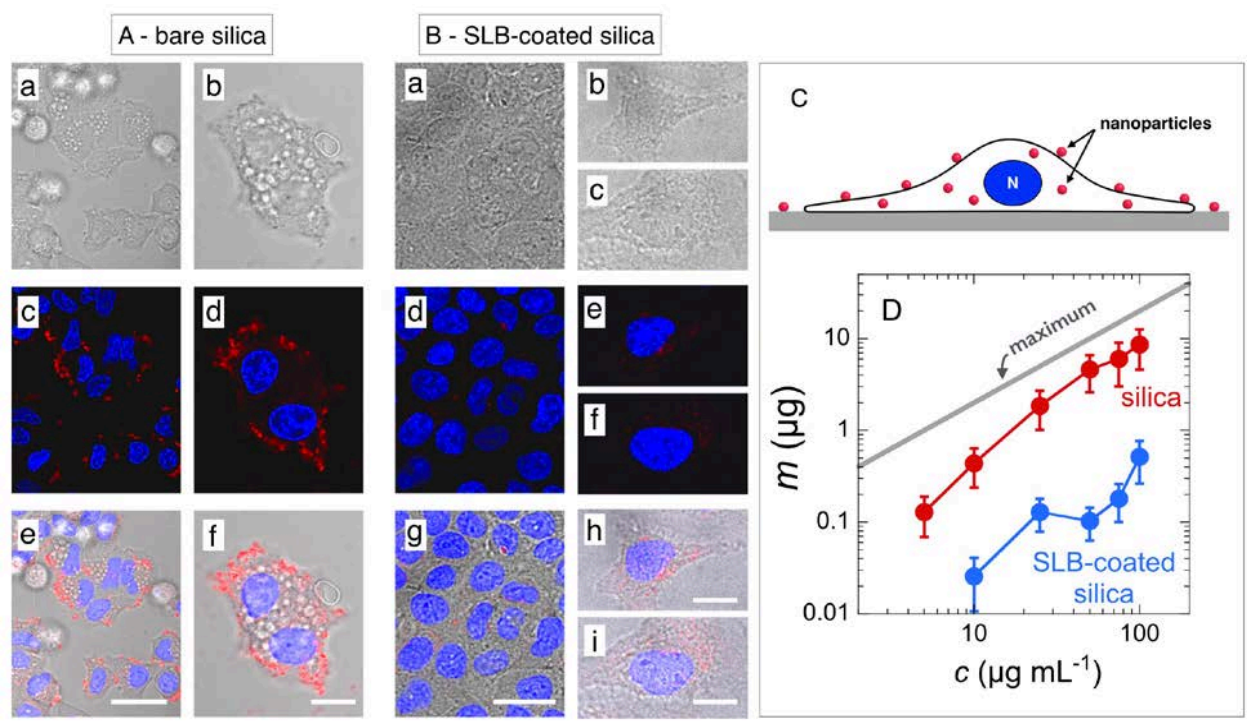

***Figure 7:*** *A) Bright field (a,b), confocal microscopy (c,d) and merge (e,f) images of A549 alveolar epithelial cells incubated with bare rhodamine-loaded fluorescent silica. In the bright field images, micron sized vacuoles are observed. The blue and red fluorescence channels correspond to DAPI and rhodamine emissions, respectively. The bars in (a,c,e) are 20 µm and in (b,d,f) 10 µm. B) Bright field (a-c), confocal microscopy (d-f) and merge (g-i) images of A549 alveolar epithelial cells with fluorescent silica coated with a 5 nm thick Curosurf® supported lipid bilayer. The bars in (a,d,g) are 20 µm and in (b,c,e,f,h,i) 10 µm. Additional confocal images are shown in Supporting Information S12. C) Schematic representation of a A549 cell with adsorbed and internalized nanoparticles. D) Mass of internalized-adsorbed silica as a function of the dose for bare aminated and SLB-coated fluorescent silica nanoparticles.*

*II. 4 - Intracellular localization*

Here we determine the experimental conditions under which the fluorescent silica particles interact with epithelial cells and where they are located in the cytoplasm. The particle localization was investigated by TEM on 70 nm thick microtomed cell sections. The exposure conditions were 4 h and silica concentration of 1 g $L^{-1}$. Figs. 8a – 8e display representative images of A549 cells incubated with bare particles. In Fig. 8a, an image of a cell suggests the presence of micrometer large intracellular vacuoles similar to those seen in optical microscopy. These vacuoles are the signature that the cytoplasm integrity was damaged and that cells are undergoing a cellular death process. On the same figure, nanoparticle aggregates are also visible beyond the cell limits (arrows).

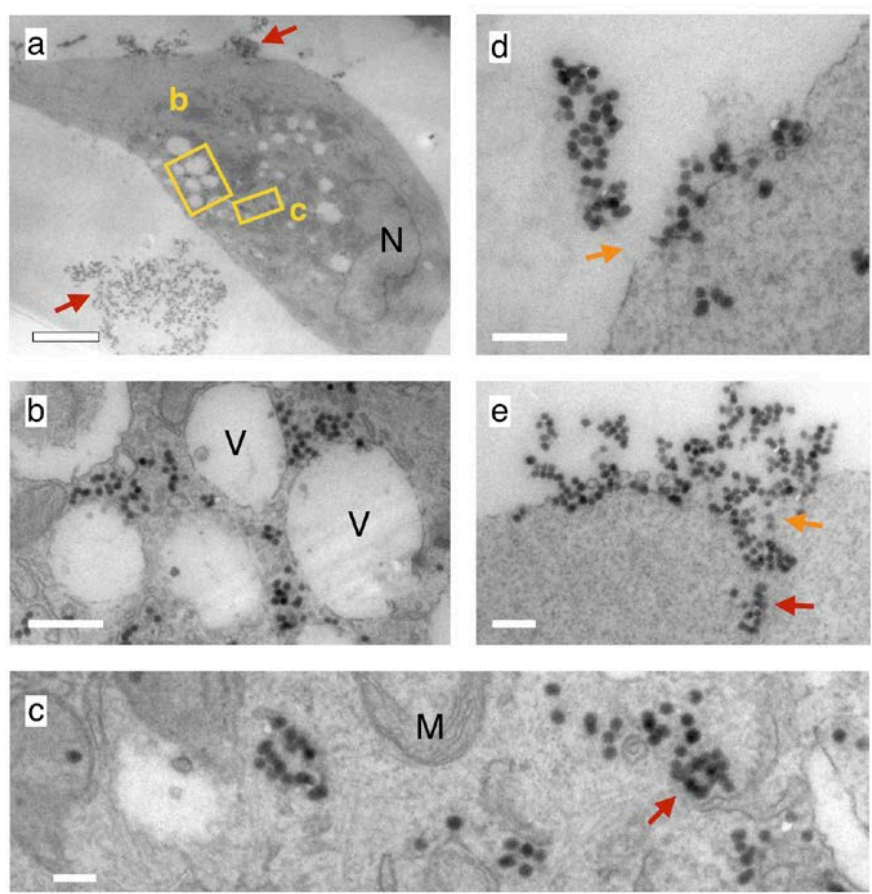

**Figure 8:** *a-e) TEM images of A549 alveolar epithelial cells incubated with rhodamine-loaded fluorescent silica particles. The red arrows in a), c) and e) highlight the presence of nanoparticle aggregates. The orange arrows in d) and e) point out to examples of disrupted cell membrane. N, V and M denote nucleus, vacuole and mitochondria respectively. Bars are 3 µm in a), 500 nm in b) and 200 nm in c-e).*

Close-up views illustrated in Figs. 8b and 8c indicate that the large vacuoles are empty, confirming the absence of co-localization seen in confocal microscopy. Moreover, every detected particle appears to be in the cytosol and not in membrane bound-compartments. In most cases, the particles are aggregated in clusters of a few units. In this assay, nanoparticles were found neither in the nucleus nor in mitochondria. Fig. 8d and 8e show examples of aggregates adsorbed at the outer lipid layer. In the first figure the membrane appears to be disrupted over 100 nm after the passage of a group of 4 particles (arrow). In the second, the particles engulf into the cytosol without deforming the membrane. These findings differ from those found with anionic or neutral particles for which internalization occurs *via* endocytosis and formation of endosomal compartments.[64-67] Pertaining to the entry mechanism, the results suggest an internalization process *via* membrane perforation, most likely induced by charge and amine group effects.[62]

Fig. 9 displays TEM images of cells treated with SLB-coated fluorescent silica in the same conditions (see also Supporting Information S13). There are here substantial changes compared to Fig. 8. The TEM images do not show any vacuole and aggregate at the outer membrane or in surrounding fluid, and the whole cell seems to contain fewer particles (Fig. 9a). This outcome is confirmed in the lower panels (Figs. 9b-d), where magnified views reveal individual particles or small clusters located only in membrane-bound compartments.[64,65,67] These compartments are identified as lysosomes, as their interiors have a electron dense lumen.[68] A few endosomes were nevertheless observed in the TEM images. With the current spatial resolution, it is not possible to conclude whether the supported layers are still present, or if the lipids have fused with the outer cellular membrane. Further studies are necessary to conclude about this point. Particles were not found in the nucleus or associated with other intracellular organelles. These results are consistent with those obtained from toxicity and internalization rate discussed previously. They are also in line with recent *in vitro* data on 50 nm silica particles exposed to cells in the presence of serum.[66] There, the protein corona ensured a weak cellular membrane adhesion and a low internalization efficiency, highlighting the role of biomolecular coating in the nanomaterial-cell interactions.

# III – Conclusion

Previous studies suggest the potential formation of hybrid structures between the pulmonary surfactant lining the alveoli and inhaled particles. Here we examine one of these interaction scenarios and study the role of supporting lipid bilayers on the fate of fluorescently labeled silica nanoparticles towards a malignant epithelial cell line. For that we focus on Curosurf®, an exogenous surfactant administered to premature babies suffering from respiratory distress syndrome. In the context of the present study, Curosurf® offers many advantages: *i)* it possesses the membrane proteins SP-B and SP-C which are essential for the air-liquid interface structuration; *ii)* its physico-chemical properties are also well controlled by the manufacturer and experiments are reproducible from batch to batch; *iii)* it also exhibits excellent temporal stability, allowing to perform extensive characterization and interaction experiments.[16] Using a combination of light scattering, zeta potential and cryo-TEM experiments, it is first shown that the Curosurf® is made of multivesicular vesicles of sizes in the range 100 nm - 5 μm.

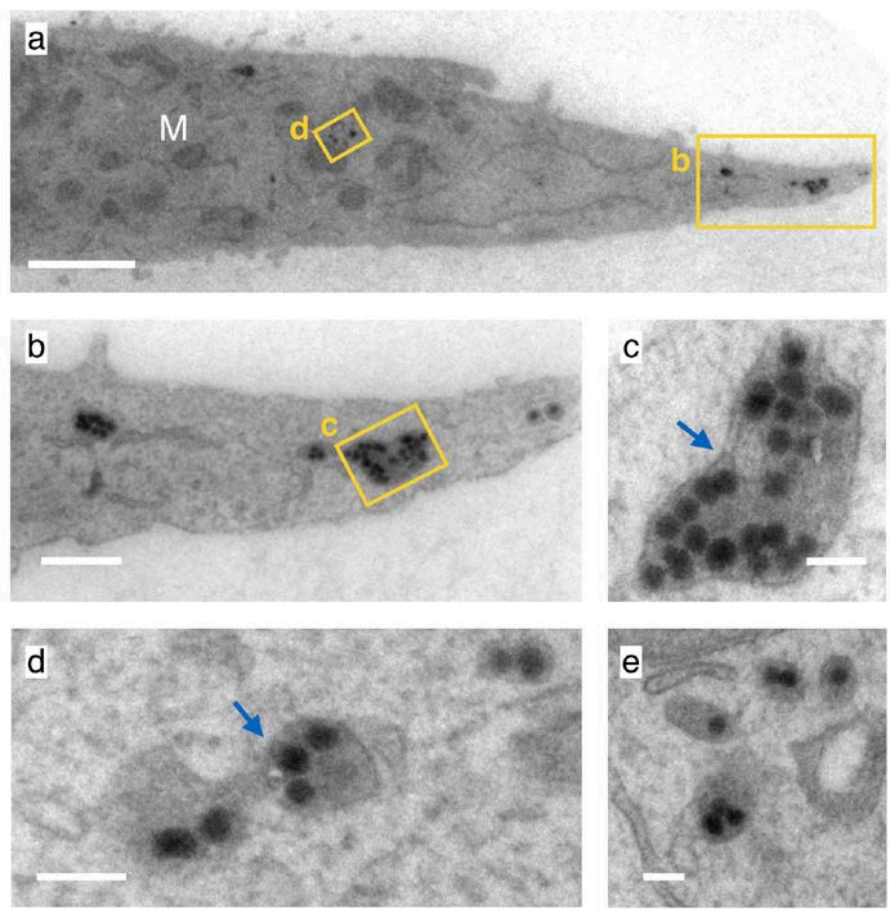

**Figure 9:** *TEM images of A549 alveolar epithelial cells incubated with SLB-coated fluorescent silica particles. The arrows are pointing out to nanoparticles enclosed in cellular compartments. The coated particle dispersion were prepared at $X_S = 2$ according to the protocol described in Section II.1. Bars are 2 μm in a), 500 nm in b) and 100 nm in c-e). Additional TEM images are shown in Supporting Information S13.*

To enhance the particle-membrane adhesion energy, the particles were made positive so that they can interact with the negatively charged Curosurf® vesicles *via* Coulombic forces. In the current work, we establish the multi-component phase diagram of these 40 nm silica particles with the pulmonary surfactant model, Curosurf®. When mixed together, strong electrostatic interactions result and lead to the formation of micrometer large aggregates. To produce SLBs, we start from such aggregates and applied sonication. The fine-tuning of the silica and vesicle specific surfaces enables to determine an optimum nanoparticle-to-vesicle surface $X_S$. Above $X_S = 2$ all particles observed upon cryo-TEM are covered with a single supported bilayer. In the process the particles reverse their charges, from positive at low $X_S$ to negative above $X_S = 2$. Instances of monolayers or multi-bilayers coating are not found using this approach. The SLB characteristics are obtained from a combination of different structural measurements, including centrifugal sedimentation, transmission and cryo-transmission electron microscopy. It is concluded that cryo-TEM provides the most accurate determination for the SLB thickness, $\delta_{SLB} = 5.2 \pm 0.6$ nm, in good agreement with that of Curosurf® membranes. The other techniques used either overestimate (like light scattering and TEM) or to underestimate (DCS) this thickness. Once formed, the SLB coated particles exhibit a robust stability in various solvents, a result that is ascribed to the strong local electrostatic interaction between the phospholipid head groups and the surface amines. To our knowledge, it is the first time that SLBs from surfactant substitute are achieved on particles below 100 nm.

In a second part, we explored the impact of surfactant SLBs on the cytotoxic potential and interactions towards A549 lung epithelial carcinoma cells. In contrast to bare particles which elicit a marked toxicity, SLB-coated silica display a rather uniform cellular responses with a cell viability remaining around 100% up to the highest dose level. It is also shown that SLB-coated silica are 20 to 50 times less internalized compared to uncoated nanoparticles. In the cytoplasm, the particles localization is also strongly coating dependent. Bare particles are massively internalized and found freely dispersed in the cytosol, whereas particles with SLB are located exclusively in endosomes and lysosomes. In conclusion we show here that supported lipid bilayers made from pulmonary surfactant and deposited on nanoparticles durably affect cellular interactions and functions *in vitro*. The current data also shed some light on potential mechanisms pertaining to the particle transport through the air-blood barrier, and confirm that lung lining fluids represent an efficient barrier against pathogens and harmful particles. The deposition and the adsorption of biologically relevant membranes on curved interfaces should also open up new ways for functionalizing inorganic surfaces.

# IV – Materials and Methods

## IV.1 – Materials

*Surfactant substitutes*

Curosurf®, also called Poractant Alfa (*Chiesi Pharmaceuticals*, Parma, Italy) is porcine minced pulmonary surfactant extract.[11] It is produced as a 80 g L$^{-1}$ phospholipid and protein suspension containing among others phosphatidylcholine lipids, phosphatidylglycerol and the membrane proteins SP-B and SP-C.[46] Curosurf® is indicated for the rescue treatment of Respiratory Distress Syndrome (RDS) in premature infants and is administered at a dose of 200 mg per kilogram. In the present study, Curosurf® is considered as a pulmonary surfactant model that offers substantial advantages: its formulation is well controled and its composition and internal structure are not susceptible to changes from one batch to another (in contrast for instance with endogenous surfactant). According to *Chiesi*, the pH of Curosurf® is adjusted with sodium bicarbonate to pH 6.2, which is close to the pH of the endogenous pulmonary surfactant.[51,52] The actual pH measured on different batches was in fact comprised between 5.5 and 6.5.[16] Curosurf® was kindly provided by Dr. Mostafa Mokhtari and his team from the neonatal service at Hospital Kremlin-Bicêtre, Val-de-Marne, France.

For the protein-free surfactant studied with the Continuous Variation Method, we used a mixture of phospholipids made from dipalmitoylphosphatidylcholine (DPPC), L-α-Phosphatidyl-DL-glycerol sodium salt from egg yolk lecithin (PG) and 2-Oleoyl-1-palmitoyl-sn-glycero-3-phospho-rac-(1-glycerol) (POPG). The lipids were initially dissolved in methanol at 10, 10 and 20 g $L^{-1}$ respectively and then mixed in proper amounts for a final weight concentration of 80% / 10% / 10% of DPPC/PG/POPG. The solvent was evaporated under low pressure at 60 °C for 30 minutes. The lipid film formed on the bottom of the flask was then rehydrated with the addition of Milli-Q water at 60 °C and agitated at atmospheric pressure for another 30 minutes. Milli-Q water was added again to finally obtain a solution at 1 g $L^{-1}$.

*Fluorescent silica nanoparticles*

The positively charged silica particles were synthetized using the Stöber synthesis.[69] Briefly, fluorescent silica seeds were prepared using rhodamine and silica precursors.[70] In a second step a non-fluorescent silica shell was grown to increase the particle size. Functionalization by amine groups was then performed, resulting in a positively charged coating.[71] Aminated silica were synthesized at 40 g $L^{-1}$ and diluted with DI-water at pH 6. The particles were characterized by light scattering and transmission electron microscopy, yielding for the hydrodynamic diameter $D_H$ = 60 nm and for the core diameter 41.2 nm. Fluorescence properties were characterized using a Cary Eclipse fluorimeter (Agilent), leading excitation and emission peaks at 572 nm and 590 nm respectively. The identity card of the fluorescent silica particles, including data from UV-Vis spectrometry, fluorimetry and light scattering is provided in Supporting Information S3.

*Chemicals*

Phosphate buffer saline (PBS1X), trypsin–EDTA, DMEM GlutaMAX$^{TM}$, DMEM without red phenol (called white DMEM in the following), fetal bovine serum (FBS) and penicillin-streptomycin were purchased from Gibco, Life Technologies. 2-(4-iodophenyl)-3-(4-nitrophenyl)-5-(2,4-disulphophenyl)-2H-tetrazolium (WST-1) was bought from Roche Diagnostics, Basel, Switzerland. Water was deionized with a Millipore Milli-Q Water system. All the products were used without further purification.

## IV.2 – Experimental techniques

*Mixing and sonication protocols*

Mixed fluorescent silica and surfactant dispersions were characterized by the surface ratio $X_S$, which denotes the specific surface of membranes associated to the vesicles divided by that developed by the particles. The nanoparticle stock solutions were prepared at 6 g $L^{-1}$ and at pH 6 in DI-water while the Curosurf® concentrations were adjusted to have the desired surface ratio. As already mentioned, the requirement on the pH was imposed by the particle and vesicle physico-chemical properties and by the fact that it corresponds to the mild acidic conditions encountered in alveolar spaces. Adjusting the pH of the stock solutions also ensures that no aggregation of nanoparticles occurs owing to the pH gap.[72] Upon mixing, interactions between particles and vesicles occurred rapidly due to their opposite surface charges.[48-50] Pertaining to the sonication, a sonicator bath (Bioblock Scientific, model 89202) working at the frequency of 35 kHz and an applied power of 55 W was used. During sonication, the temperature was maintained between 30 °C and 40 °C. For light scattering and zeta potential experiments, silica-surfactant dispersions were diluted between 0.1 and 1 g $L^{-1}$ in particles to reduce absorption.

*Static and Dynamic Light scattering*

The scattered intensity $I_S$ and the hydrodynamic diameter $D_H$ were determined using a NanoZS Zetasizer (Malvern Instruments). The second-order autocorrelation function is analyzed using the cumulant and CONTIN algorithms to determine the average diffusion coefficient $D_C$ of the scatterers. Hydrodynamic diameter is then calculated according to the Stokes-Einstein relation $D_H = k_B T/3\pi\eta D_C$ where $k_B$ is the Boltzmann constant, $T$ the temperature and $\eta$ the solvent viscosity. Measurements were performed in triplicate at 25 °C and 37 °C after an equilibration time of 120 s.

*Electrophoretic mobility and zeta potential*

Laser Doppler velocimetry using the phase analysis light scattering mode and detection at an angle of 16° was used to carry out the electrokinetic measurements of electrophoretic mobility and zeta potential with the Zetasizer Nano ZS equipment (Malvern Instruments, UK). Zeta potential was measured after a 120 s equilibration at 25 °C.

*Transmission electron microscopy (TEM and cryo-TEM)*

TEM imaging was performed with a Tecnai 12 operating at 80 kV equipped with a 1k×1k Keen View camera. A 3 μL drop of silica-surfactant dispersion (concentration 0.05 g L$^{-1}$) was deposited on holey-carbon coated 300 mesh copper grids (Neyco). The sample was stained during 30 s with uranyl acetate at 0.5 wt. %. Uranyl acetate provides a dark labeling of the phospholipid heads due to the high affinity of the electron dense ion uranyl to carboxyl groups. Grids were let to dry at room temperature in the dark for 20 min. For cryo-TEM, few microliters of surfactant (concentration 5 g L$^{-1}$ in DI-water) or silica–surfactant dispersions (concentration 0.5 g L$^{-1}$ in white DMEM) were deposited on a lacey carbon coated 200 mesh (Ted Pella Inc.). The drop was blotted with a filter paper using a FEI Vitrobot$^{TM}$ freeze plunger. The grid was then quenched rapidly in liquid ethane to avoid crystallization and later cooled with liquid nitrogen. The membrane was then transferred into the vacuum column of a JEOL 1400 TEM microscope (120 kV) where it was maintained at liquid nitrogen temperature thanks to a cryo-holder (Gatan). The magnification was comprised between 3000× and 40000×, and images were recorded with an 2k×2k Ultrascan camera (Gatan). TEM and cryo-TEM images were digitized and treated by the ImageJ software and plugins (http://rsbweb.nih.gov/ij/).

*Differential Centrifugal Sedimentation*

Particle size distributions were measured using a DCS disc centrifuge DC24000 UHR (CPS Instruments Inc.). Sucrose solutions at 8 and 24 wt. % were freshly prepared in DI-water, and mixed in order to have solutions at 22, 20, 18, 16, 14, 12 and 10 wt. %. 1.6 mL of solutions were filled successively in nine steps, from 24 to 8 wt. % into the disc rotating at a speed of 21000 rpm. Calibration was performed using poly(vinyl chloride) particles (0.239 μm, Analytik Ltd.) as calibration standard before each measurement. As particles are centrifuged through the sucrose gradient, they pass through the detector beam and the absorbance increases. The absorbance *versus* time data is then translated into the size number distribution of the particles. For the mixed silica-surfactant samples, 100 μL of nanoparticles (concentration 1.5 g L$^{-1}$) were injected into the disc centrifuge, and each sample was analyzed in triplicate to verify data reproducibility. The mean of these three results was calculated for further data analysis.

*Optical and confocal microscopy*

For confocal microscopy, samples were examined with an inverted wide-field microscope (Olympus IX81) equipped with a spinning disk module (Yokogawa CSU-X1), with an oil immersion objective (60×, NA 1.42) and a EMCCD camera (Andor iXon 897). Z-stacks of wide-field fluorescent images were acquired using a piezo at 0.2 μm increments. For the DAPI channel, 405 nm laser-line excitation and a 465 nm-centered emission filter were used. For the red channel, the excitation line was 488 nm and the emission filter was centered at 590 nm. Addition confocal experiments were made at the Plateforme ImagoSeine (Institute Jacques Monot, Paris, France) using a LSM 710 (Zeiss) microscope equipped with an oil immersion 40× objective.

# IV.3 – Cellular biology

*A549 epithelial cell culture*

Adenocarcinomic human alveolar epithelial cells A549 (ATCC reference CCL-185[TM]) were grown in T75-flasks in DMEM supplemented with 10% FBS and 1% penicillin-streptomycin. Exponentially growing cultures were maintained in a humidified atmosphere of 5% $CO_2$ and 95% air at 37 °C. When the cell confluence reached 80%, cell cultures were passaged using trypsin–EDTA.

*WST-1 Toxicity assay*

Cells were seeded in a 96-well plate at 20000 cells per well in white DMEM with serum. After 24 h, cells were rinsed with PBS and incubated for another 24 h at 37 °C with 200 μL of a fluorescent silica dispersion (in white DMEM). Cells were rinsed with PBS and incubated for 30 min with 100 μL of WST-1 per well (dilution 1:20 in white DMEM). The assay is based upon the reduction by cellular dehydrogenase of the achromatic tetrazolium salt WST-1 to the dark yellow colored soluble formazan. The generation of formazan was measured at 450 nm in a microplate reader (ELx808, Biotek Instruments, Winooski, United States) against a blank containing culture media and WST-1 and data were corrected from the absorbance at 630 nm. Positive controls were performed with hydrogen peroxide ($H_2O_2$) and the non-interference between nanoparticles, cells and WST-1 was assessed treating half of the cultures with Triton (2% in white DMEM, 15 min of incubation at 37 °C) according to Vietti *et al.*[73] Data are represented as means ± standards deviations ($n = 3$) and are analyzed with the GraphPad Prims 7 software using variance analysis (one-way ANOVA) followed by Dunnett's test with p = 0.0332, 0.0021, 0.002 and 0.0001 (one to four stars respectively).

*Cell preparation for confocal microscopy*

A549 epithelial cells were seeded on coverslips in 6-well chamber at 200000 cells per well in white DMEM with serum. After 48 h, cells were incubated 4 h at 37 °C with fluorescent silica particles (1 mL at 1 g $L^{-1}$ in white DMEM), then rinsed with PBS and finally fixed during 20 min at 37°C with 700 μL of a 4% paraformaldehyde solution (J61899, Alfa Aesar) in PBS. After rinsing, the nuclei were fluorescently stained with DAPI (Aldrich) by incubation for 20 min at room temperature in dark conditions (dilution 1:1000 in PBS). The cells were thereafter washed three times with PBS to remove the dye in excess. The coverslips were then taken out of the wells and mounted onto a Gene Frame® adhesive system.

*Quantitative determination of adsorption/internalization efficiency*

A549 epithelial cells were seeded in a 96-well plate at 10000 cells per well in white DMEM with serum. After 48 h, the cells were rinsed with PBS and incubated for 4 h at 37 °C with fluorescent silica nanoparticles, with a dose from 5 to 100 μg mL$^{-1}$. After rinsing with white DMEM, fluorescence was recorded using a Flexstation 3 (Molecular Device, $\lambda_{ex}$ = 533 nm, $\lambda_{em}$ = 590 nm, cut-off 570 nm). The fluorescence intensity was read from the bottom of the plate. Fluorescence intensities were converted into mass of particles thanks to a calibration curve performed in white DMEM.

*TEM on cells*

A549 epithelial cells were seeded in T25-flasks. After 48 h (confluence 80%), cells were rinsed with PBS and incubated for 4 h at 37 °C with 5 mL of silica or silica-Curosurf® solutions (concentration of 1 g L$^{-1}$ in white DMEM). Cells were washed with PBS and fixed with a glutaraldehyde - PFA buffer (1% - 2.5% respectively in PBS) for 1 h at room temperature. Fixed cells were further washed with PBS, scratched and centrifuged. The pellets were then postfixed during 1 h in a 1% osmium tetroxide solution reduced by a 1.5% potassium ferrocyanide solution. After several washes, the samples were dehydrated by addition of ethanol – propylene oxide mixtures and wrapped up in epoxy resin. 70 nm-thick sections were cut with an ultramicrotome (LEICA, Ultracut UC6) and picked up on copper/formvar carbon grids. They were then stained for 7 min with 4% uranyl acetate and for 7 min with 0.2% lead citrate.

# Acknowledgments

We thank Mélody Merle, Mostafa Mokhtari, Evdokia Oikonomou, Nicolas Tsapis for fruitful discussions. ANR (Agence Nationale de la Recherche) and CGI (Commissariat à l'Investissement d'Avenir) are gratefully acknowledged for their financial support of this work through Labex SEAM (Science and Engineering for Advanced Materials and devices) ANR 11 LABX 086, ANR 11 IDEX 05 02. We acknowledge the ImagoSeine facility (Jacques Monod Institute, Paris, France), and the France BioImaging infrastructure supported by the French National Research Agency (ANR-10-INSB-04, « Investments for the future »). This research was supported in part by the Agence Nationale de la Recherche under the contract ANR-13-BS08-0015 (PANORAMA) and ANR-12-CHEX-0011 (PULMONANO) and by Solvay.

## TOC Figure

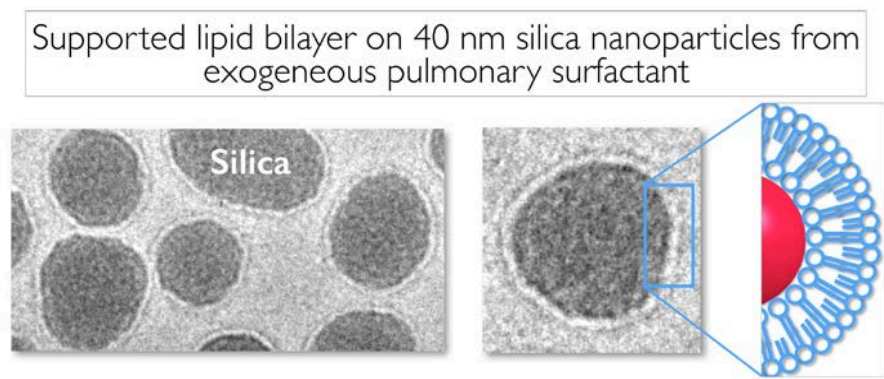